\documentstyle[sprocl,epsf]{article}

\def\be{\begin{equation}}
\def\ee{\end{equation}}
\def\bea{\begin{eqnarray}}
\def\eea{\end{eqnarray}}

\newcommand{\gsim}{\;\rlap{\lower 3.5 pt \hbox{$\mathchar \sim$}} \raise 1pt
 \hbox {$>$}\;}
\newcommand{\lsim}{\;\rlap{\lower 3.5 pt \hbox{$\mathchar \sim$}} \raise 1pt
 \hbox {$<$}\;}

\newcommand{\logtwos}{L_{tW}}

\begin{document}

\hfill BUTP-98/26

\hfill hep-ph/9811342

\vspace{2em}

\title{COMPLETELY AUTOMATED CALCULATIONS OF MULTI-LOOP DIAGRAMS}

\author{
M. STEINHAUSER
}

\address{
  Institut f\"ur Theoretische Physik, Universit\"at Bern,\\ 
  Sidlerstr. 5, CH-3012 Berne\\
  E-mail: Matthias.Steinhauser@itp.unibe.ch
}

\maketitle
\abstracts{
The computation of higher order processes very often involves
a large number of diagrams. In addition, it is in general
not possible to solve the occurring integrals explicitly and
expansions in small quantities have to be performed. This makes
it necessary to automate the calculations as much as possible.
A program package will be described which generates automatically
the Feynman diagrams, manipulates the expressions in the desired 
way and performs the computation.
As a physical application ${\cal O}(\alpha \alpha_s)$ corrections
to the decay rate of the $Z$ boson into bottom quarks are discussed.
}


\section{Introduction and motivation}

The impressive experimental precision reached so far mainly at LEP,
SLC and TEVATRON has made it necessary to increase also the effort from the
theoretical side. A crucial role plays thereby the computation of multi-loop
diagrams in order to evaluate quantum corrections to the different
observables. 

Higher order corrections are mostly accompanied with a large number of
diagrams. Very often it is a tedious job to do the bookkeeping and not to
forget some relevant contributions. Especially
when the number exceeds the order of a few 
hundred the generation should be passed to the computer. A further
major task is the computation of the integrals. In most cases an exact
solution is by far not possible and one has to rely on approximations. 
One possibility it to compute the integrals 
(at least partly) numerically. Another
promising attempt is the use of asymptotic expansions which is applicable as
soon as a certain hierarchy exist between the mass scales involved in the
process.
Well-defined prescriptions provide rules which specify the actions on the
individual diagrams. In general each diagram
generates several subgraphs which
further increase the complexity of the calculation.
Thus it is desired to automate the asymptotic expansion procedures.
Of course, also the very computation of the single terms needs to be done by
the computer as the size of intermediate expression become rather
large.

In the next section a possible solution to the problems addressed above is
presented by means of the package {\tt GEFICOM}.
In Section~\ref{sec:appl} its application to the
computation of ${\cal O}(\alpha\alpha_s)$ corrections to the decay of the $Z$
boson into bottom quarks is discussed.


\section{{\tt GEFICOM}: A package for generation and computation of
  Feynman diagrams}

\begin{figure}[b]
\begin{center}
\leavevmode
\epsfxsize=8cm
\epsffile[77 200 520 760]{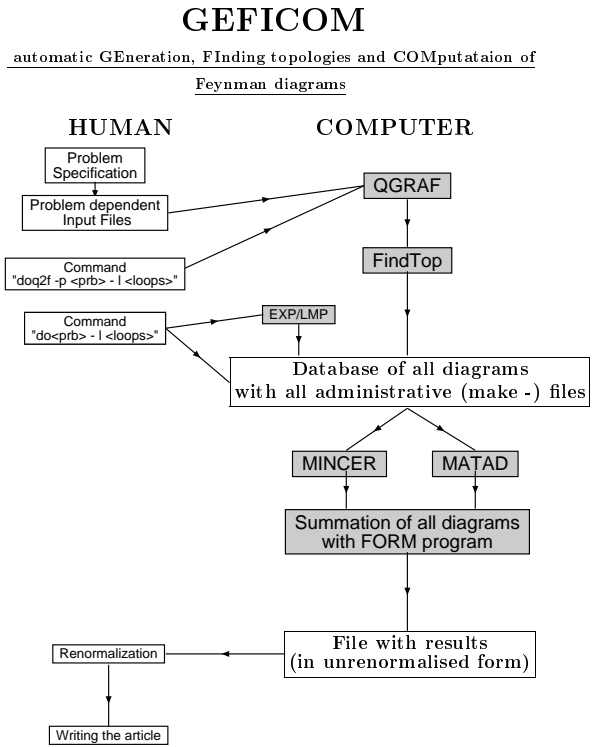}
\caption{The structure of {\tt GEFICOM}.
}
\label{fig:fchart}
\end{center}
\end{figure}

The automation of the computation of Feynman diagrams can be divided into
different steps.
In this section we will discuss them on the example of 
{\tt GEFICOM}~\cite{geficom}.
A flowchart is pictured in Fig.~\ref{fig:fchart}.

The user has to provide a few simple files specifying, e.g., the process, the
particle content and the allowed vertices. Then, in a first step, the
graphs
contributing to the considered process have to be generated
where the amplitudes at least need to contain information about the
involved particles and the momenta flowing through each propagator.
Inside {\tt GEFICOM} the {\tt FORTRAN} program 
{\tt QGRAF}~\cite{Nog93} is used which has the advantage of being
quite fast. Moreover different output formats are available one of which
contains all the desired information on the diagrams.

The Feynman diagrams are classified according to their topology
which becomes especially important during the calculation of the integrals.
This
information is, however, not provided by {\tt QGRAF} and has to be determined
from its output.
For a human being this would be very easy. For a computer, on the other
side, this is a quite non-trivial
task and the momentum distribution has to be used
in a clever way in order to extract the topology.
This is done by {\tt FindTop} which is part of a
{\tt Mathematica} program essentially inserting the Feynman rules
and transforming the {\tt QGRAF}
output into {\tt FORM}~\cite{form} notation.
Additionally administrative files are generated which rule
the computation, sum up the diagrams etc.
One finally ends up with a huge database containing all relevant files.

The very computation of the diagrams has to be initiated by the user.
It is based on two {\tt FORM} packages, {\tt MINCER}~\cite{MINCER} and
{\tt MATAD}~\cite{Ste96},
being able to deal with massless two-point, respectively,
massive bubble diagrams up to three loops.
Before, however, the amplitudes are passed to these packages there is the
possibility to apply asymptotic expansions namely the
hard-mass and large-momentum
procedure providing consistent prescriptions on how to get expansions
for large internal masses, respectively, large external momenta.
They work on a diagram-by-diagram basis and make well-defined rules
available which
determine the subgraphs to be extracted from the original diagram~\cite{hmp}.
The subgraphs have to be expanded in their small quantities before any
momentum integration is performed. Thus significant simplifications are
obtained and very often the integrals can be written as products of lower
order ones at the price of --- sometimes significantly ---
increasing their number.
The automation of these procedures has been performed in the packages
{\tt LMP}~\cite{Har:diss} and {\tt EXP}~\cite{Sei:dipl}.
They apply the
asymptotic-expansion procedures and express the initial diagram
as products of single-scale integrals which can be treated by 
{\tt MINCER} and {\tt MATAD}.

A very powerful tool for the computation of the diagrams is
implemented into {\tt MINCER} and {\tt MATAD}, the
so-called integration-by-parts technique. Up to
now it has been systematically applied up to the three-loop
level~\cite{CheTka81,Bro92,LapRem96}.
The general idea is quite simple: The integration-by-parts method exploits
the fact that the momentum integrals in $D$ dimensions are finite. Thus the
surface terms are zero:
\begin{eqnarray}
  \int{\rm d}^D p {\partial\over \partial p^\mu} f(p,\ldots) &=& 0\,.
\label{eq:ip}
\end{eqnarray}
The function $f$ in general is a product of scalar propagators belonging to
the considered loop. Performing the derivatives in Eq.~(\ref{eq:ip})
explicitly leads to a sum of several terms which has to vanish identically.
Combining different equations of this type, which are obtained by 
varying $f(p,\ldots)$ and choosing different loop momenta, leads to 
relations --- so-called recurrence relations --- where a complicated integral 
is expressed as a sum of simpler ones.
The repeated application of recurrence relations finally leads to
a sum simple integrals, which can be reduced to integrals of lower loop order,
and only a few, so-called, master integrals which require a hard
calculation.

At the three-loop level
the described strategy has first been applied to massless two-point
integrals~\cite{CheTka81} and has later on been extended to massive
bubble diagrams~\cite{Bro92}.
Also the corresponding master integrals are available~\cite{Bro98}.
These two sets of
recurrence relations were implemented into {\tt MINCER} and {\tt MATAD},
respectively, and we will see
in the next section that they provide when combined with the
asymptotic expansions mentioned above a very powerful tool with a large
field of application.


\section{\label{sec:appl}Application}

The properties of the $Z$ boson have been measured to a very high
accuracy --- sometimes even at the permille level. Also the hadronic width of
the $Z$ boson, $\Gamma_{\rm had}$, is known to an accuracy of roughly $0.1$\%.
An important contribution to $\Gamma_{\rm had}$ is provided by the partial
rate into bottom quarks as already at one-loop order the top quark
enters as a virtual particle giving rise to enhanced corrections
proportional to $G_F M_t^2$. The full electroweak corrections are known since
quite some time~\cite{AkhBarRie86,BeeHol88} and also the leading terms of 
${\cal O}(\alpha_sG_F M_t^2)$~\cite{FleJegRacTar92} and
${\cal O}(\alpha\alpha_s\ln M_t^2/M_W^2)$~\cite{KwiSte95}
are available since already a few
years. Actually it turned out that the logarithmic contribution is relatively
large as compared to the term quadratic in $M_t$
although $\ln(M_t^2/M_W^2)$ is by far less enhanced than $M_t^2/M_W^2$.
Thus the question arises whether the $M_t$-independent coefficient and
the power-suppressed terms in the expansion lead to a significant change in
the numerical prediction.

\begin{figure}[t]
\begin{center}
\leavevmode
\epsfxsize=10.5cm
\epsffile[105 558 503 717]{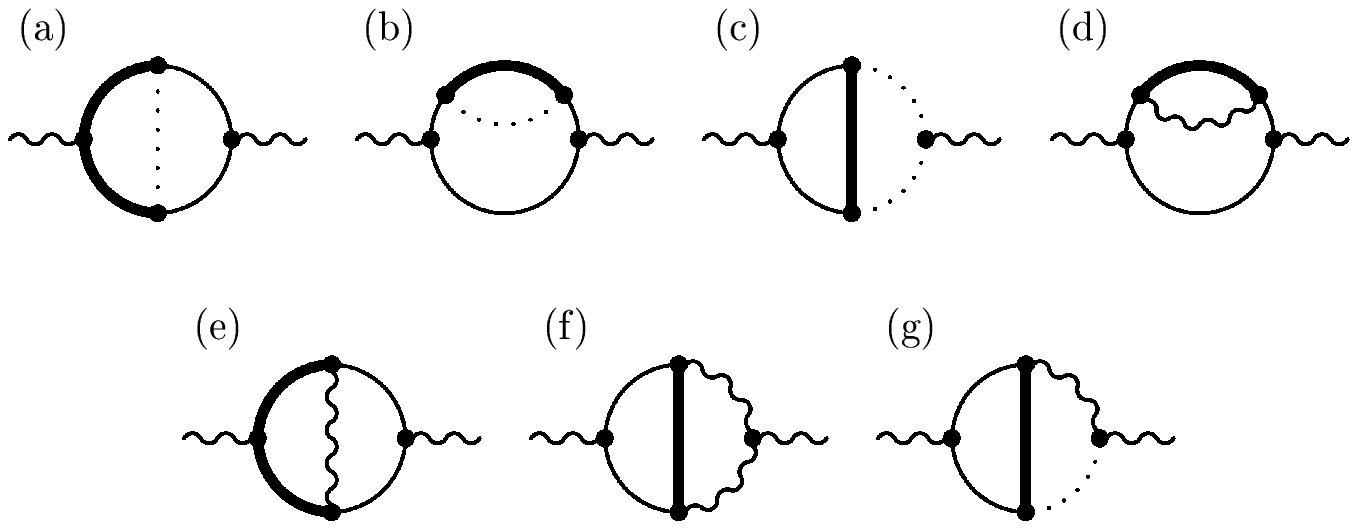}
\caption{Diagrams contributing to $\Gamma(Z\to b\bar{b})$ and containing 
  a virtual top quark.
  Thin lines correspond to bottom quarks, thick
  lines to top quarks, dotted lines to Goldstone bosons and inner wavy
  lines represent $W$ bosons.}
\label{fig:zbbdias}
\end{center}
\end{figure}

In the approach chosen in~\cite{HarSeiSte97}
the $Z$ boson propagator was considered whose imaginary part directly leads to
the decay width. The diagrams contributing at the two-loop level are
shown in Fig.~\ref{fig:zbbdias}. A gluon has to be added in all possible ways
in order to obtain the graphs leading to corrections of order
$\alpha\alpha_s$.

For vanishing bottom quark mass up to four scales may appear inside a single
diagram: $M_t^2$, $M_W^2$, $\xi_W M_W^2$ and $q^2$ where $q$ is the external
momentum to be identified with $M_Z^2$ at the end.
$\xi_W$ is the electroweak gauge parameter.
In order to be able to apply the hard-mass procedure a certain hierarchy has
to be fixed. The assumption $M_t^2\gg (M_W^2,q^2=M_Z^2)$ is clearly justified.
Furthermore the diagram of Fig.~\ref{fig:zbbdias}$(f)$
makes it necessary to choose $M_W^2\gg q^2$ in order to avoid contributions
to the imaginary part arising from cutting lines involving $W$ bosons.
At first sight this
looks seemingly inadequate. However, a closer look to the corresponding
diagrams shows that actually $(2M_W)^2$, respectively, $(M_t+M_W)^2$ has to be
compared with $q^2$ which justifies the above inequality.
This is true, since the real emission
of a $W$ boson is always accompanied by another $W$ boson or a top quark.
The notation $M_W^2\gg q^2$ is only formal, telling the hard-mass
procedure which subgraphs have to be selected.

Concerning the fourth scale, $\xi_W M_W^2$,
there is some freedom for the choice at which place
in the inequality chain it is inserted. Only $q^2\gg\xi_W M_W^2$
is not allowed as then again unwanted imaginary parts would be generated.
The other three possibilities are allowed and must lead to
identical final results as the
$\xi_W$ dependence drops out at the very end. In intermediate steps,
however, the expressions which have to be evaluated are different.
Thus the different choices provide quite strong checks on the calculation.

At two-loop level seven diagrams have to be considered. Their
calculation would still
be feasible by hand. At three loops, however, 69 diagrams contribute
and a computation by hand is very painful
especially if higher order terms in the $1/M_t$ expansion are considered.
The hard-mass procedure
applied to the 69 initial three-loop diagrams
results in 234 subdiagrams which have to be expanded in their
small quantities.
In~\cite{HarSeiSte97} the package {\tt EXP} written in {\tt Fortran~90}
was used which automatically applies 
the hard-mass procedure. It can be called form 
{\tt GEFICOM} setting appropriate options in the
initial files to be provided by the user.

Since we are interested in the virtual effect of the top quark
we consider in the following the difference of the partial widths
of the $Z$ boson into bottom and down quarks:
\begin{eqnarray}
\delta\Gamma_{b-d}^W&=&
  \delta\Gamma_b^{0,W} - \delta\Gamma_d^{0,W}\,,
\label{eq:gambd}
\end{eqnarray}
where $\delta\Gamma_d^{0,W}$ is the contribution 
from the diagrams involving a $W$ boson to the partial decay rate
$\Gamma(Z\to d\bar{d})$~\cite{CzaKue96}.
$\delta\Gamma_b^{0,W}$ is obtained from the diagrams in Fig.~\ref{fig:zbbdias}
and the corresponding ones at three-loop order.
The zero indicates that next to the vertex diagrams no additional
counterterms had been introduced as they would drop out in the difference.
Note that besides the pole parts also the dependence on $\xi_W$
drops out in Eq.~(\ref{eq:gambd}). The final result reads in numerical
form~\cite{HarSeiSte97}:
\begin{eqnarray}
\delta\Gamma^W_{b-d} &\approx& \Gamma^0
  {1\over s_\theta^2}{\alpha\over \pi}\times
\nonumber\\&&
\bigg\{
       - 0.11\,{M_t^2\over M_W^2}
       + 0.71 
       - 0.31\,\logtwos 
       + \left(
             0.36
           - 0.89\,\logtwos
          \right)\,{M_W^2\over M_t^2} 
\nonumber\\&&\mbox{}
       + \left(
           - 0.24
           - 0.97\,\logtwos
          \right)\,\left({M_W^2\over M_t^2}\right)^{2} 
       + \left(
           - 0.78
           - 0.43\,\logtwos
          \right)\,\left({M_W^2\over M_t^2}\right)^{3} 
\nonumber\\&&\mbox{}
       + {\alpha_s\over \pi}\,\bigg[
             0.24\,{M_t^2\over M_W^2}
           + 1.21
           - 0.32\,\logtwos 
           + \left(
                 1.40
               - 1.99\,\logtwos
              \right)\,{M_W^2\over M_t^2} 
\nonumber\\&&\mbox{}\hspace{1em}
           + \left(
                 0.37
               - 2.99\,\logtwos 
               + 0.08\,\logtwos^2
              \right)\,\left({M_W^2\over M_t^2}\right)^{2} 
\nonumber\\&&\mbox{}\hspace{1em}
           + \left(
               - 1.08
               - 2.64\,\logtwos 
               + 0.17\,\logtwos^2
              \right)\,\left({M_W^2\over M_t^2}\right)^{3} 
          \bigg] 
\bigg\}
+ {\cal O} \left( \left( {M_W^2\over M_t^2} \right)^4 \right)
\nonumber\\
&\approx&
    \Gamma^0 {1\over s^2_\theta}
    {\alpha\over \pi}
  \bigg\{ - 0.50
  + (0.71 -0.48)+ (0.08 - 0.29) + (-0.01 - 0.07) 
  \nonumber\\[.5em]&&\mbox{\hspace{1em}}
+ (-0.007 - 0.006)
  + {\alpha_s\over \pi} \bigg[ 1.16 + (1.21 - 0.49) + (0.30 - 0.65) 
\nonumber\\&&\mbox{\hspace{1em}}
+ (0.02 - 0.21 + 0.01)
+ (-0.01 - 0.04
  + 0.004) \bigg] \bigg\}
\nonumber\\&\approx&\mbox{\hspace{0em}}
\Gamma^0 {1\over s^2_\theta} {\alpha\over \pi} \bigg\{- 0.50 - 0.07 +
{\alpha_s\over \pi} \bigg[ 1.16 + 0.13 \bigg]\bigg\}\,,
\label{eqdelWnum}
\end{eqnarray}
with $\Gamma^0 = N_cM_Z\alpha/(12s_\theta^2c_\theta^2$), $s_\theta =
\sin\theta_W$, $\theta_W$ being the weak mixing angle,
$c^2_\theta=1-s^2_\theta$ and
$\logtwos = \ln(M_t^2/M_W^2)$.
$M_t$ is the on-shell top mass. 
The numbers after the second equality sign correspond to
successively increasing orders in $1/M_t^2$, where the brackets collect
the corresponding constant, $\ln M_t$ and, if present, $\ln^2
M_t$ terms. The numbers after the third equality sign represent the
leading $M_t^2$ term and the sum of the subleading ones where the ${\cal
  O}(\alpha)$ and ${\cal O}(\alpha\alpha_s)$ results are displayed
separately. 

One observes that for the realistic values $M_Z=91.91$~GeV, $M_t=175$~GeV 
and $s^2_\theta=0.223$, which were used in Eq.~(\ref{eqdelWnum}),
the constant at next-to-leading order dominates over the 
$\ln M_t^2$ term known before.
It is remarkable
that at one-loop level the corrections arising from the $1/M_t^2$ terms are 
of similar size than the one from next-to-leading order, however,
the signs are different. The higher order corrections in $1/M_t$
are smaller, which means that effectively only the leading $M_t^2$ term
remains.

Proceeding to two loops the situation is similar: Starting at
${\cal O}(1/M_t^2)$ the sign is opposite as compared to the leading
terms and a large cancellation takes place. Here, the $1/M_t^4$
term is still comparable with the $1/M_t^2$ contribution.
The $1/M_t^6$ term, however, is considerably smaller
which suggests that the presented terms should provide a reasonable
approximation to the full result.
Comparison of this expansion of the one-loop terms to the
exact result of \cite{BeeHol88} shows agreement up to $0.01\%$
which also gives quite some confidence in the ${\cal O}(\alpha\alpha_s)$
contribution.

For $\alpha = 1/129$ and
$\alpha_s(M_Z)=0.120$ we get:
\begin{eqnarray}
  \delta\Gamma_{b-d}^W &=& (-5.69 - 0.79 + 0.50 + 0.06)\mbox{ MeV}
\,\,=\,\, -5.92\mbox{ MeV} \,.
\label{eqdelwnum}
\end{eqnarray}
The first two numbers in Eq.~(\ref{eqdelwnum}) correspond to the 
${\cal O}(\alpha)$, the second two to the ${\cal O}(\alpha\alpha_s)$
corrections.
Each of these contributions is again separated into the $M_t^2$ terms and the
sum of the subleading ones.


\section*{Acknowledgments}

It is a pleasure to thank K.G.~Chetyrkin, R.~Harlander, J.H.~K\"uhn, and
T.~Seidensticker for fruitful collaborations and the organizers of RADCOR~98
for the nice atmosphere during the conference.


\end{document}